\documentclass[letterpaper,journal]{IEEEtran}
\usepackage{amssymb,amsfonts}
\usepackage[cmex10]{amsmath}
\usepackage{algorithmic}
\usepackage{algorithm}
\usepackage{array}
\usepackage{textcomp}
\usepackage{stfloats}
\usepackage{url}
\usepackage{verbatim}
\usepackage{graphicx}
\usepackage{cite}
\usepackage{amsthm}
\usepackage{booktabs}

\newcolumntype{L}[1]{ >{\raggedright\arraybackslash}m{#1}} 
\usepackage{bm}
\ifCLASSOPTIONcompsoc
  \usepackage[caption=false,font=normalsize,labelfont=sf,textfont=sf]{subfig}
\else
  \usepackage[caption=false,font=footnotesize]{subfig}
\fi
\usepackage[table,xcdraw]{xcolor}
\usepackage{diagbox}
\usepackage{makecell}
\usepackage{multirow}

\begin{document}
\title{ASTRA: Toward Agentic AI for Intelligent Device-Network-Cloud Synergy in Next-Generation Mobile Communication}
\author{
	\IEEEauthorblockN{
		Yalong Guo\textsuperscript{1,2}, 
		Jinbo Tan\textsuperscript{1}, 
        Ying Wang\textsuperscript{2},
        Fan Zhang\textsuperscript{2},
		Jintao Wang\textsuperscript{1}, 
		and Changyong Pan\textsuperscript{1}\\
	}
	\IEEEauthorblockA{
		\textit{\textsuperscript{1}Department of Electronic Engineering}, Tsinghua University, Beijing, China \\
		\textit{\textsuperscript{1}State Key Laboratory of Space Network and Communications}, Tsinghua University, Beijing, China \\
        \textit{\textsuperscript{2}2012 Theory Lab}, Huawei Technologies Co. Ltd, Beijing, China \\
		Emails: guo-yl24@mails.tsinghua.edu.cn, tanjingbo@tsinghua.edu.cn, wangying188@huawei.com, \\ zhang.fan2@huawei.com, wangjintao@tsinghua.edu.cn, pcy@tsinghua.edu.cn
	}
}
\maketitle

\begin{abstract}
    The evolution toward next-generation mobile communication systems demands intelligence-native networks capable of autonomously adapting to user intent, yet the prevailing 3GPP protocol-driven device-network-cloud (DNC) architecture imposes three structural bottlenecks: protocol-constrained decision spaces confining optimization to predefined parameter subsets, cascaded information asymmetry from lossy interface compression that strips semantic context and causes intent miscalibration, and inherently reactive coordination mechanisms that trigger actions only after performance degradation. This paper proposes an autonomous agentic AI paradigm named Agentic Synergy for Telecommunication Resource Autonomy (ASTRA), which introduces a three-tier agent layer, including device agent, network agent, and cloud agent, decoupling network intelligence from the underlying hardware infrastructure. These agents collaborate through bidirectional semantic channels, including semantic intent messages, capability abstraction messages, global directives, and peer coordination, executing a six-phase cycle of perceive, reason and predict, communicate, decide, act, and learn that transforms reactive protocol-driven operations into proactive, intent-calibrated optimization over the full decision space. Validated through system-level simulations in two representative scenarios, ASTRA achieves a 13.1\% average throughput gain in dense-crowd cell selection by redistributing UEs from congested cells via semantic load exchange, and an 18.2\% passive handover reduction in high-speed mobility through predictive trajectory-aware coordination, providing initial evidence that the proposed agentic framework accesses solution regions structurally inaccessible under protocol-constrained architectures.
\end{abstract}

\begin{IEEEkeywords}
Agentic AI, Device-network-cloud synergy, Next-generation mobile communication, Large language model, Intent-driven networking
\end{IEEEkeywords}

\section{Introduction}
\label{sec:introduction}

    \subsection{Research Background}

    \IEEEPARstart{T}{he} evolution toward next-generation mobile communication systems represents a fundamental paradigm shift from connectivity-centric to intelligence-native networks, where ubiquitous intelligence enables networks to autonomously adapt to user intent, proactively provision resources, and orchestrate services across heterogeneous computational tiers~\cite{wang_road_2023, you_towards_2020, giordani_toward_2020}. Central to this vision is the device-network-cloud continuum, which serves as a hierarchical architecture that distributes intelligence and computational resources across end devices, edge infrastructure, and centralized cloud data centers~\cite{liu_joint_2023, wang_cooperative_2024, peng_6g_2025}. The DNC continuum was originally engineered under the design philosophy of layered protocol decomposition, standardized interfaces, and predefined parameter spaces. This approach proved highly effective for the ``connect everything'' era of 4G/5G but now encounters fundamental bottlenecks in the ``intelligence-native'' era of 6G~\cite{cui_overview_2025}.

    These bottlenecks stem from the prevailing 3GPP protocol-driven architecture, which organizes DNC cooperation through a layered protocol stack where each tier operates within predefined parameter spaces and exchanges information via structured signaling~\cite{ghosh_5g_2019, tataria_6g_2021}. Specifically, three interrelated bottlenecks arise: (1)~protocol-constrained decision spaces confining optimization to a predefined subset of the achievable performance region; (2)~information asymmetry from cascaded interface compression that strips away semantic context; and (3)~reactive coordination mechanisms preventing anticipatory action. These bottlenecks share a common root: layered decomposition fragments information and standardization rigidifies parameters, which together force reactive decision-making~\cite{antonopoulos_agile-6g_2025, polese_understanding_2023}.
    
    Recent breakthroughs in large language models (LLMs) and autonomous agent systems have catalyzed the emergence of agentic AI~\cite{zhou_large_2025, yang_decision-making_2026}, i.e., systems characterized by autonomous decision-making, goal-directed behavior, and adaptive learning, which is now being integrated into the ecosystem of next-generation mobile communications~\cite{abou_ali_agentic_2025, xiao_toward_2025}. These agentic AI-empowered capabilities, such as autonomous network management, intent-driven resource orchestration, and cross-domain fault diagnosis, hold immense potential to address the structural bottlenecks of the current DNC architecture by introducing an intelligence layer that operates above the protocol stack, enabling agents to reason across layer boundaries, exchange semantic intent, and negotiate coordination strategies beyond protocol-constrained search spaces~\cite{long_survey_2025, qayyum_llm-driven_2025}.

    Driven by the need to mitigate these structural bottlenecks, we propose a DNC continuum paradigm named ASTRA, which decouples network intelligence from infrastructure. The proposed paradigm introduces a new agent layer between the physical layer and the function layer, enabling the deployment of agents at the device, the edge, and the cloud respectively. Then, the interactive protocols and data flows among these agents are defined by leveraging semantic transmission. Finally, we develop the collaborative operation workflow under the proposed framework, to realize proactive response optimization based on agentic AI. This architecture decouples intelligence from the underlying hardware infrastructure, and through bidirectional semantic channel-based collaboration, enables cloud-edge-device cooperation over the full decision space, thereby yielding tangible performance improvements. Validated through dense-crowd cell selection and high-speed mobility optimization, ASTRA achieves 13.1\% average throughput gain and 18.2\% passive handover reduction respectively.

    \subsection{Summary of Contributions}

    The central thesis of this work is that making communication systems agentic—embedding autonomous, reasoning-capable agents across the DNC continuum—is the essential paradigm shift for next-generation mobile networks. ASTRA provides the architectural vision and conceptual framework for this shift; realizing a production-grade agentic communication system remains a long-term research endeavor.

    This paper makes three principal contributions:

    \begin{itemize}
        \item \textbf{A Three-Tier Agentic DNC Architecture.} We propose ASTRA, a novel paradigm that introduces an agent layer between the physical and function layers, deploying Device Agents (DA), Network Agents (NA), and Cloud Agents (CA) across the DNC continuum. This architecture decouples network intelligence from the underlying hardware infrastructure, thereby expanding the feasible decision space from the protocol-constrained subset $\mathcal{X}^{\mathrm{3GPP}}$ to the full space $\mathcal{X}$ and enabling cross-layer optimization strategies structurally inaccessible under the prevailing 3GPP protocol-driven architecture.

        \item \textbf{Bidirectional Semantic Channels and Proactive Coordination Cycle.} We design four semantic-level interaction protocols---semantic intent messages (DA$\to$NA), aggregated telemetry reports (NA$\to$CA), global directives (CA$\to$NA), and peer coordination messages (NA$\leftrightarrow$NA)---that replace the cascaded, lossy interface compression of traditional architectures with rich bidirectional context exchange. Building on these channels, we formalize a six-phase collaborative cycle (Perceive, Reason \& Predict, Communicate, Decide, Act, Learn) that transforms reactive protocol-driven operations into proactive, intent-calibrated coordination over heterogeneous perception scopes.

        \item \textbf{System-Level Validation in Representative Scenarios.} We validate ASTRA through system-level simulations in two canonical 6G scenarios: dense-crowd cell selection and high-speed mobility optimization. Results demonstrate a 13.1\% average throughput gain via semantic load-aware UE redistribution and an 18.2\% reduction in passive handovers via predictive trajectory-aware coordination, confirming that the agentic framework accesses solution regions structurally inaccessible under protocol-constrained architectures.
    \end{itemize}

    \subsection{Paper Organization}

    The remainder of this paper is organized as follows. Section~\ref{sec:tradition} characterizes the traditional 3GPP-driven DNC architecture and provides a rigorous formulation of its three structural bottlenecks---protocol-constrained decision spaces, information asymmetry, and reactive coordination---establishing the baseline against which the proposed paradigm is evaluated. Section~\ref{sec:paradigm} presents the ASTRA framework, detailing the three-tier agent layer, the four classes of bidirectional semantic channels, and the three design principles that directly counteract the identified bottlenecks. Section~\ref{sec:workflow} formalizes the six-phase agentic coordination cycle and provides a taxonomy of supported application domains, with concrete instantiations for dense-crowd cell selection and high-speed mobility optimization. Section~\ref{sec:performance} validates ASTRA through system-level simulations in both scenarios, reporting quantitative performance gains over the traditional protocol-driven baseline. Section~\ref{sec:conclusion} concludes the paper.

\section{Traditional DNC Architecture and Limitations}
\label{sec:tradition}

    \subsection{Traditional DNC Architecture}
    
    Before presenting the proposed agentic paradigm, we characterize the traditional DNC architecture to establish the structural baseline against which our framework is evaluated. The traditional architecture, governed by 3GPP specifications~\cite{li_6g_2022}, organizes the DNC continuum into three functionally isolated tiers interconnected through standardized interfaces.

    \textbf{Device Tier.} The UE executes application workloads and interacts with the network through the Uu interface. The UE's role in system optimization is limited to two functions: (i)~performing radio measurements (RSRP, RSRQ, SINR) and reporting them via RRC measurement reports upon network-configured event triggers, and (ii)~transmitting Buffer Status Reports (BSR) and Power Headroom Reports (PHR) to inform uplink scheduling. The UE possesses no mechanism to convey application-layer semantics, user intent, or Quality of Experience (QoE) perception to the network---its communication with the infrastructure is restricted to a predefined set of physical-layer indicators.

    \textbf{Network Tier.} The Radio Access Network (RAN), comprising the gNodeB's distributed unit (DU), centralized unit (CU), and associated Mobile Edge Computing (MEC) infrastructure, performs real-time radio resource management. The DU executes MAC-layer scheduling on a per-TTI basis using instantaneous Channel State Information (CSI); the CU manages RRC-layer procedures including connection management, handover control, and bearer configuration; and the MEC platform provides edge computational resources for latency-sensitive offloading. Decision-making at this tier is confined to protocol-defined parameter spaces: scheduling weights from a fixed algorithm set, handover thresholds within 3GPP-specified ranges, and QoS enforcement through standardized 5QI-to-resource mappings. Inter-cell coordination is limited to the Xn interface, which exchanges load indications and handover signaling but lacks semantic content about user intent or application context.

    \textbf{Cloud Tier.} The 5G Core (5GC) and centralized cloud infrastructure handle session management (SMF), user plane routing (UPF), policy control (PCF), and large-scale data analytics (NWDAF). The cloud tier operates at the coarsest temporal granularity (minutes to hours) and relies on aggregated KPI telemetry for network-wide optimization. Its control over the RAN is mediated through policy rules disseminated via the PCF$\to$gNB path, which translate high-level SLA requirements into per-slice resource quotas and QoS profiles. The cloud has no direct access to real-time radio conditions or device-level context; its decisions are necessarily based on delayed, aggregated, and compressed information.

    \subsection{Limitations of Traditional Architecture}
    \label{subsec:limitation}

    To precisely delineate the research gap, we elaborate on the three structural limitations inherent to the 3GPP-driven DNC architecture. These limitations cannot be resolved through incremental improvements; they require a paradigm shift in how intelligence is organized across the DNC continuum.

        \subsubsection{Protocol-Constrained Decision Spaces}

        All operational decision variables in traditional DNC systems, including handover thresholds, scheduling weights, QoS mappings, and offloading policies, are confined to parameter spaces predefined by 3GPP specifications, leaving vast regions of superior solutions structurally inaccessible. We define the full decision space as:
        \begin{equation}
        \mathcal{X} = \mathcal{X}_{\mathrm{radio}} \times \mathcal{X}_{\mathrm{compute}} \times \mathcal{X}_{\mathrm{control}}.
        \end{equation}
        This space integrates the radio, computing, and control domains into a unified decision-making framework. The optimal coordination strategy solves:
        \begin{equation}
        \mathbf{x}^{\star} = \arg\max_{\mathbf{x} \in \mathcal{X}} U(\mathbf{x}),
        \end{equation}
        with optimal utility $U^{\star} = U(\mathbf{x}^{\star})$.

\begin{figure}[!t]
            \centering
            \includegraphics[width=0.38\textwidth]{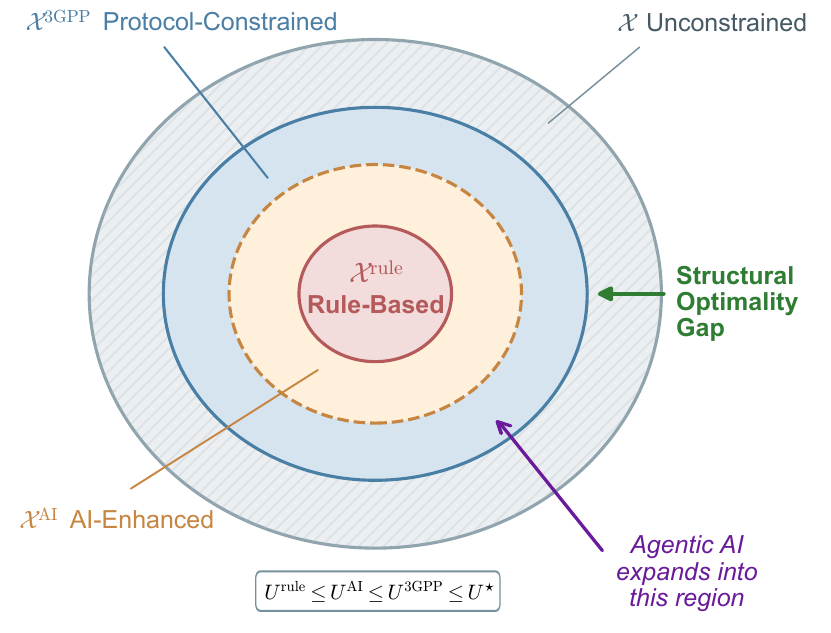}
            \caption{The nested decision-space hierarchy.}
            \label{fig:decision_space}
        \end{figure}

        Within the 3GPP framework, three progressively larger feasible regions exist, as shown in Fig.~\ref{fig:decision_space}. Rule-based methods operate with fixed heuristics (e.g., A3-event handover with static hysteresis, round-robin scheduling), restricting decisions to $\mathcal{X}^{\mathrm{rule}} \subset \mathcal{X}$. AI-enhanced methods~\cite{lee_intelligent_2022, zhang_deep_2020, eisen_optimal_2020, zheng_large_2026, she_efficient_2024} replace these heuristics with learned policies exploring broader parameter ranges, enlarging the feasible region to $\mathcal{X}^{\mathrm{AI}} \supseteq \mathcal{X}^{\mathrm{rule}}$. Let $\mathcal{X}^{\mathrm{3GPP}}$ denote the maximal feasible region under the standardized protocol framework. Notably, both rule-based and AI-enhanced strategies are fully constrained and compliant with the standardized 3GPP protocol framework. This yields:
        \begin{equation}
            \mathcal{X}^{\mathrm{rule}} \subseteq \mathcal{X}^{\mathrm{AI}} \subseteq \mathcal{X}^{\mathrm{3GPP}} \subseteq \mathcal{X},
            \label{eq:decision_space}
        \end{equation}
        and the corresponding utility hierarchy:
        \begin{equation}
            U^{\mathrm{rule}} \leq U^{\mathrm{AI}} \leq U^{\mathrm{3GPP}} \leq U^{\star}.
            \label{eq:utility_chain}
        \end{equation}
        The inequality $U^{\mathrm{3GPP}} \leq U^{\star}$ is the central observation: the optimality gap originates from the architecture's structural constraints, not from suboptimal parameter selection within it.

        \subsubsection{Information Asymmetry and Intent Miscalibration}

        The cascaded interface architecture (Uu, F1, E1, NG, N6) imposes fixed information schemas that destroy contextual information through lossy compression. Let $\mathbf{I}_u(t)$ denote the full intent state of user $u$. At the Uu interface, this is reduced to:
        \begin{equation}
            \mathbf{I}_{u}(t) \xrightarrow{\text{Uu}} \mathbf{m}_{u}(t) = \bigl[\mathrm{RSRP}, \mathrm{RSRQ}, \mathrm{SINR}, \mathrm{CQI}\bigr]^\top,
            \label{eq:info_compression}
        \end{equation}
        with loss $\mathcal{L}_{\mathrm{Uu}} = H(\mathbf{I}_u \mid \mathbf{m}_u) > 0$. At the NG interface, per-UE QoS requirements collapse into a finite set of standardized 5QI classes. By the data processing inequality, the cumulative loss grows monotonically:
        \begin{equation}
            \mathcal{L}_{\mathrm{total}} = H(\mathbf{I}_u) - I(\mathbf{I}_u; \mathbf{o}_{\mathrm{DN}}).
            \label{eq:total_loss}
        \end{equation}
        This asymmetry is bidirectional: the UE has no visibility into cell-level resource utilization or peer scheduling, causing application-layer adaptation to operate on stale information.

        \subsubsection{Reactive Rather Than Proactive Coordination}

        The traditional framework relies exclusively on reactive triggering: actions are initiated only after performance degradation has been detected and reported. Consider handover: the A3 event fires when the neighboring cell's RSRP exceeds the serving cell's RSRP by $\theta_{\mathrm{hys}}$ for duration $T_{\mathrm{TTT}}$, yielding total latency:
        \begin{equation}
            T_{\mathrm{HO}}^{\mathrm{total}} = T_{\mathrm{degrade}} + T_{\mathrm{TTT}} + T_{\mathrm{prep}} + T_{\mathrm{exec}} + T_{\mathrm{complete}},
            \label{eq:ho_latency}
        \end{equation}
        where $T_{\mathrm{HO}}^{\mathrm{total}}$ denotes total handover latency, $T_{\mathrm{degrade}}$ is the link degradation duration, $T_{\mathrm{TTT}}$ is the time-to-trigger timer, $T_{\mathrm{prep}}$ is handover preparation latency, $T_{\mathrm{exec}}$ is execution latency, and $T_{\mathrm{complete}}$ is link re-establishment latency. Throughout $[0, t_{\mathrm{trigger}} + T_{\mathrm{prep}}]$, the UE operates on a degrading link with no compensatory action.

        This reactive pattern pervades the framework: KPI reporting at coarse granularity, threshold alarms that fire only after violations persist, slice reconfiguration on administrative timescales, and per-TTI scheduling without temporal foresight. Next-generation applications (autonomous driving, remote surgery, industrial automation) demand anticipatory provisioning that prevents failures rather than reacting to them.

\section{ASTRA: A Proposed DNC Synergy Paradigm}
\label{sec:paradigm}

To address the three structural limitations identified in Section~\ref{sec:tradition}, this section proposes ASTRA: a paradigm that decouples intelligence from infrastructure and enables intent-calibrated, proactive coordination across the DNC continuum. ASTRA is built upon three design principles: intelligence-infrastructure decoupling, semantic-level information exchange, and proactive coordination, which directly counteract the identified bottlenecks, shifting the system's operating point from \(U^{\mathrm{3GPP}}\) toward the theoretical optimum \(U^{\star}\).

\begin{figure*}[!t]
    \centering
    \includegraphics[width=0.85\textwidth]{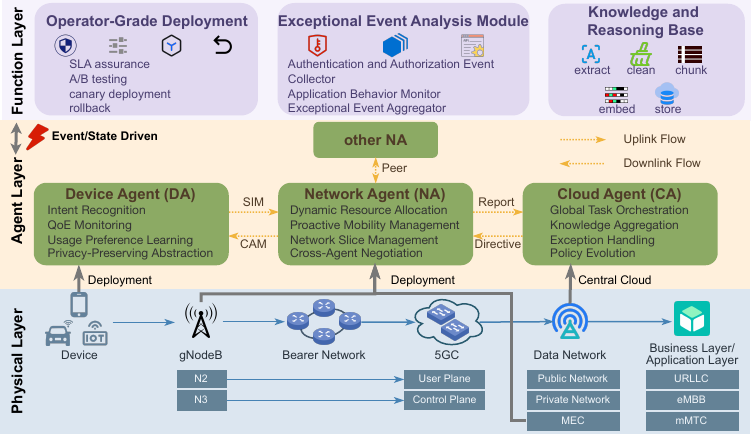}
    \caption{Conceptual Architecture of the Proposed ASTRA Paradigm. The architecture comprises three vertically integrated layers: a standards-compliant physical layer providing the execution substrate, an agent layer hosting device agents, network agents, and cloud agents, and a function layer supplying cross-cutting capabilities. Solid arrows denote physical-layer data flows; dashed arrows denote semantic-level agent interactions.}
    \label{fig:framework}
\end{figure*}

    \subsection{Proposed Agentic AI-Based DNC Framework}

    The ASTRA framework introduces an agent layer that abstracts decision-making from the underlying infrastructure, decoupling network intelligence from the physical deployment topology. As depicted in Fig.~\ref{fig:framework}, the framework is organized into three vertically integrated layers.
    
    A distinguishing feature of the proposed architecture is the explicit modeling of heterogeneous perception scopes across agent classes. The DA perceives a local/private scope encompassing device-level telemetry, application-layer context, and user behavioral patterns, which is rich in semantic content but narrow in spatial coverage. The NA perceives a cell-level scope spanning radio resource utilization across all served UEs, channel state information, cell load dynamics, and MEC resource availability, providing moderate spatial coverage with real-time granularity. The CA perceives a global/historical scope aggregating network-wide KPI trends, cross-region traffic patterns, long-term policy performance, and knowledge from all NAs, offering the broadest coverage but with inherent latency.

    The partial overlap of these perception scopes creates natural information-sharing zones where agents exchange complementary observations, collectively constructing a more complete system view than any single agent could achieve independently. This design transforms the unidirectional information flow of traditional architectures into a collaborative perception fusion process that systematically reduces the mutual information loss $\mathcal{L}_{\mathrm{total}}$.

        \subsubsection{Physical Layer}

        The Physical Layer retains the standard 5G end-to-end data path. It provides the execution substrate for radio transmission, baseband processing, core network routing, and edge/cloud computation, upon which agent decisions are enacted. Critically, this layer remains fully standards-compliant; the agentic paradigm operates above the protocol stack rather than replacing it, ensuring incremental deployability in operational networks.

        The Physical Layer exposes two types of interfaces to the Agent Layer: (i)~Telemetry interfaces: Agents observe the physical-layer state through standardized and extended measurement reporting, including radio-level indicators (RSRP, RSRQ, SINR, CQI), transport-level metrics (throughput, latency, jitter), and computational-level status (CPU/GPU utilization, memory, thermal state). (ii)~Control APIs: Agents enact decisions through configuration interfaces that map semantic-level directives to physical-layer parameter adjustments, including scheduling configuration, power control, beamforming weights, handover commands, and slice resource quotas.
        
        This dual-interface design ensures that the Agent Layer can both perceive and act upon the Physical Layer.

        \subsubsection{Agent Layer}

                \begin{table*}[!t]
            \centering
            \caption{Agent Competency Mapping Across the DNC Continuum}
            \label{tab:competency_mapping}
            \renewcommand{\arraystretch}{1.3}
            \small
            \begin{tabular}{|p{2.0cm}|p{1.8cm}|p{2.5cm}|p{9.0cm}|}
                \hline
                \textbf{Agent Class} & \textbf{Deployment} & \textbf{Perception Scope} & \textbf{Primary Competencies} \\
                \hline
                Device Agent & UE chipset & Local / Private & Current status sensing, Communication experience monitoring, Usage preference learning, Intent recognition \\
                \hline
                Network Agent & gNB / MEC & Cell-level & Dynamic resource allocation, Route optimization and traffic steering, Network slice management, Proactive mobility management \\
                \hline
                Cloud Agent & Central cloud & Global / Historical & Global task orchestration, Knowledge sharing and aggregation, Exception handling and conflict resolution, Long-term policy evolution \\
                \hline
            \end{tabular}
        \end{table*}

        The Agent Layer is the intelligence core of the proposed framework, hosting three classes of autonomous agents with differentiated roles, perception scopes, and operational time scales. Unlike the rigid, protocol-driven control logic of traditional architectures, these agents possess autonomous perception, reasoning, and action capabilities.

            \paragraph{Device Agent}

            The device agent is instantiated on the UE and serves as the user-facing intelligence endpoint. Its primary role is to bridge the gap between raw user behavior and structured semantic intent.

            The DA maintains an augmented state that extends the traditional UE state vector:
            \begin{equation}
                \mathbf{s}_u^{\mathrm{DA}}(t) = \bigl[\, \mathbf{s}_u^{\mathrm{D}}(t),\; \mathbf{e}_u(t),\; \mathbf{z}_u(t),\; \boldsymbol{\phi}_u(t) \,\bigr],
                \label{eq:da_state}
            \end{equation}
            where $\mathbf{s}_u^{\mathrm{D}}(t)$ is the conventional device state (radio measurements, battery, GPS, thermal status), $\mathbf{e}_u(t) \in \mathbb{R}^{K}$ is the QoE time-series profile capturing $K$ quality metrics over a sliding window, $\mathbf{z}_u(t) \in \mathbb{R}^{d_z}$ is the learned user preference embedding encoding implicit quality-energy trade-off preferences, and $\boldsymbol{\phi}_u(t) \in \mathbb{R}^{d_\phi}$ is the semantic intent descriptor.

            The DA's core competencies include:
            \begin{itemize}
                \item \textbf{Current Status Sensing:} Continuous monitoring and active fusion of multi-modal sensor data $\mathbf{s}_u^{\mathrm{D}}(t)$ (signal strength, battery, GPS, motion vectors, thermal status) to construct a holistic device context representation.
                \item \textbf{Communication Experience Monitoring:} Tracking real-time QoE metrics as time-series profiles $\mathbf{e}_u(t)$ rather than instantaneous snapshots, enabling trend detection and proactive degradation prediction before performance violations materialize.
                \item \textbf{Usage Preference Learning:} Building a personalized user model $\mathbf{z}_u(t)$ through continuous observation of interaction patterns, thereby capturing implicit preferences invisible to the network under traditional architectures.
                \item \textbf{Intent Recognition:} Translating observed behavior and application context into the structured semantic intent descriptor $\boldsymbol{\phi}_u(t)$, encoding task structure, criticality level, delay sensitivity, loss tolerance, and predicted resource trajectory.
            \end{itemize}

            \paragraph{Network Agent}

            The network agent is deployed on the gNB and MEC infrastructure and serves as the central coordination hub of the agentic paradigm, bridging device-level intent with cloud-level policy through real-time resource orchestration.

            The NA maintains an augmented state:
            \begin{equation}
                \mathbf{s}_n^{\mathrm{NA}}(t) = \bigl[\, \mathbf{s}_n^{\mathrm{N}}(t),\; \{\boldsymbol{\phi}_u(t)\}_{u \in \mathcal{U}_n},\; \boldsymbol{\psi}_n(t),\; \boldsymbol{\mu}_n(t) \,\bigr],
                \label{eq:na_state}
            \end{equation}
            where $\mathbf{s}_n^{\mathrm{N}}(t)$ is the conventional gNB state, $\{\boldsymbol{\phi}_u(t)\}_{u \in \mathcal{U}_n}$ are the semantic intent descriptors received from all served DAs, $\boldsymbol{\psi}_n(t)$ encapsulates the NA's predictive model state (channel prediction, load forecast, mobility prediction), and $\boldsymbol{\mu}_n(t)$ represents the current negotiation state with neighboring NAs and the CA.

            The NA's core competencies include:
            \begin{itemize}
                \item \textbf{Dynamic Resource Allocation:} Extending the traditional DU scheduler by incorporating semantic intent from DAs into the scheduling objective, jointly optimizing over an expanded action space encompassing spectrum allocation, power control, beamforming configuration, and MEC task placement.
                \item \textbf{Route Optimization and Traffic Steering:} Managing UPF selection and traffic routing decisions across multiple paths (direct backhaul, MEC local breakout, inter-gNB forwarding) based on real-time load conditions and predicted traffic evolution.
                \item \textbf{Network Slice Management:} Dynamically creating, reconfiguring, and releasing network slices in response to changing demand patterns, adapting resource quotas in real time based on aggregated intent from served DAs.
                \item \textbf{Proactive Mobility Management:} Transforming the reactive A3-event handover mechanism into a proactive, intent-aware mobility strategy by combining DA-provided mobility predictions with radio environment maps, substantially reducing handover latency.
            \end{itemize}

            \paragraph{Cloud Agent}

            The cloud agent resides in the central cloud infrastructure and provides global intelligence, long-term policy optimization, and system-wide coordination. It operates at the longest time scale but with the broadest information scope.

            The CA maintains a global state:
            \begin{equation}
                \mathbf{s}^{\mathrm{CA}}(t) = \bigl[\, \mathbf{s}^{\mathrm{C}}(t),\; \mathcal{K}(t),\; \boldsymbol{\Omega}(t) \,\bigr],
                \label{eq:ca_state}
            \end{equation}
            where $\mathbf{s}^{\mathrm{C}}(t)$ is the conventional cloud/core state, $\mathcal{K}(t)$ denotes the global knowledge base encompassing both internal operational knowledge and external domain knowledge (3GPP specifications, vendor configuration templates, semantic ontologies), and $\boldsymbol{\Omega}(t)$ represents the system-wide conflict resolution state.

            The CA's core competencies include:
            \begin{itemize}
                \item \textbf{Global Task Orchestration:} Managing cross-region task distribution and workload balancing across geographically distributed NAs based on global demand patterns and resource availability.
                \item \textbf{Knowledge Sharing and Aggregation:} Maintaining a centralized knowledge repository through federated learning, performing global model aggregation, identifying transferable knowledge across deployment scenarios, and maintaining version-controlled policy repositories.
                \item \textbf{Exception Handling and Conflict Resolution:} Serving as the arbiter when NAs encounter anomalies or inter-agent conflicts that cannot be resolved locally, applying global context to disambiguate conflicting local decisions.
                \item \textbf{Long-term Policy Evolution:} Leveraging the global/historical perception scope to identify long-term trends, seasonal patterns, and structural shifts, evolving global policies through a form of meta-learning across the agent system.
            \end{itemize}

        Table~\ref{tab:competency_mapping} summarizes the mapping between agent classes, their deployment locations, perception scopes, and primary competencies. This hierarchical design ensures that decisions are made at the appropriate temporal granularity and information scope.

        \subsubsection{Function Layer}

        The function layer provides three cross-cutting functional modules that support the full lifecycle of the agent system. Unlike the Agent Layer, which embodies what the agents decide, the Function Layer governs how agents learn, recover, and evolve over time. (i)~Operator-Grade Deployment: Carrier-level SLA assurance, multi-vendor adaptation, A/B testing, canary deployment, and rollback mechanisms for production environments. This module ensures that the agentic paradigm meets the reliability and availability requirements of operational carrier networks. (ii)~Exception Handling: Proactive anomaly detection, self-healing protocols, and conflict resolution mechanisms. When an agent detects distribution drift, anomalous inference results, or inter-agent negotiation deadlocks, it invokes this module to trigger diagnostic routines and corrective actions. (iii)~Knowledge and Reasoning Base: Maintenance of both internal operational knowledge (historical decisions, learned policies, performance records) and external domain knowledge (3GPP specifications, vendor configuration templates, semantic ontologies). The knowledge base serves as a shared memory that enables agents to leverage accumulated operational experience.

        The three-layer architecture is motivated by three design principles that directly address the structural bottlenecks identified in Section~\ref{subsec:limitation}:

            \paragraph{Intelligence--Infrastructure Decoupling}
            By separating the agent reasoning plane from the physical execution plane, the framework enables optimization over the full decision space $\mathcal{X}$ rather than the protocol-constrained subset $\mathcal{X}^{\mathrm{3GPP}}$. Agents can explore novel coordination strategies, including joint cross-layer scheduling, semantic-aware handover, and intent-driven slice reconfiguration, that are structurally inaccessible to protocol-embedded control logic. This directly addresses the protocol-constrained decision space bottleneck.

            \paragraph{Semantic-Level Information Exchange}
            The Agent Layer replaces the cascaded, lossy interface compression of traditional architectures with bidirectional semantic channels that carry intent descriptors, capability abstractions, and policy updates. This rich cross-layer state sharing minimizes the cumulative mutual information loss $\mathcal{L}_{\mathrm{total}}$ and enables each agent to construct a near-sufficient statistic for its decision scope. This directly addresses the information asymmetry and intent miscalibration bottleneck.

            \paragraph{Proactive, Anticipatory Coordination}
            Equipped with predictive models and cross-layer semantic context, agents transition from reactive triggering (waiting for performance degradation) to proactive action (anticipating and preventing degradation). Device Agents predict user mobility and intent evolution; Network Agents forecast cell load and channel conditions; Cloud Agents identify long-term demand trends. This directly addresses the reactive coordination bottleneck.

    \subsection{Interactive Workflow and Data Flow}
    \label{subsec:workflow}

    This subsection formalizes the interaction protocols between agent classes, organized by the directionality of data flows: uplink, downlink, and peer-to-peer. For each direction, we define the message formats, triggering conditions, and the role of each flow in addressing the structural limitations of traditional DNC architectures.

        \subsubsection{Uplink Flow: DA$\to$NA$\to$CA}

        The uplink flow carries semantic intent, local observations, and learning updates from the network edge toward the cloud, replacing the traditional unidirectional, lossy measurement reporting chain with a structured, multi-granularity information pathway.

        \paragraph{DA$\to$NA: Semantic Intent Upload}
        The DA transmits a Semantic Intent Message (SIM) to its serving NA:
        \begin{equation}
            \mathrm{SIM}_u(t) = \bigl\{\, \boldsymbol{\phi}_u(t),\; \mathbf{g}_u(t),\; \Delta\mathbf{e}_u(t) \,\bigr\},
            \label{eq:sim}
        \end{equation}
        where $\boldsymbol{\phi}_u(t)$ is the semantic intent descriptor encoding task type, criticality, delay/loss sensitivity, and predicted resource trajectory; $\mathbf{g}_u(t)$ is a task graph representing the computational structure and dependencies of the user's current workload; and $\Delta\mathbf{e}_u(t)$ is the QoE feedback delta capturing the change in experience metrics since the last report. Unlike the traditional measurement report that conveys only radio-level indicators, the SIM conveys what the user needs and why, preserving the semantic context that is destroyed by the cascaded interface compression of the 3GPP framework.

        \paragraph{NA$\to$CA: Aggregated Telemetry and Learning Updates}
        The NA delivers aggregated network status and exception information to the CA via a standardized periodic report message.
        \begin{equation}
            \mathrm{Report}_{n}(t) = \bigl\{\, \mathbf{KPI}_n(t),\; \mathcal{E}_n(t) \,\bigr\},
            \label{eq:na_report}
        \end{equation}
        where $\mathbf{KPI}_n(t)$ is the aggregated cell-level performance telemetry (throughput distribution, latency percentiles, handover success rates, slice utilization), and $\mathcal{E}_n(t)$ is the set of exception reports (anomalies, inter-agent conflicts, SLA violations) requiring cloud-level arbitration.

        The NA$\to$CA report serves two functions: (i)~providing the CA with the cell-level context needed for global task orchestration and policy evolution; and (ii)~escalating exceptions that exceed the NA's local resolution capability. The reporting period is adapted to network dynamics, becoming shorter during high-volatility periods and longer during stable operation, to balance communication cost against decision quality.

        \subsubsection{Downlink Flow: CA$\to$NA$\to$DA}

        The downlink flow carries global intelligence, coordination policies, and resource commitments from the cloud toward the network edge, replacing the traditional top-down, coarse-grained policy dissemination chain with a structured, semantically enriched pathway that closes the bidirectional information loop opened by the uplink flow.

        \paragraph{CA$\to$NA: Global Directives and Policy Dissemination}
        The CA transmits a Global Directive Message to each NA under its coordination scope:
        \begin{equation}
            \mathrm{Directive}_{n}(t) = \bigl\{\, \boldsymbol{\pi}_n(t),\; \hat{\boldsymbol{\rho}}_n(t),\; \Delta\mathcal{K}_n(t),\; \boldsymbol{\omega}_n(t) \,\bigr\},
            \label{eq:ca_directive}
        \end{equation}
        where $\boldsymbol{\pi}_n(t)$ is the updated policy parameter vector (e.g., optimized weighting coefficients for cell selection, slice resource quotas, scheduling priorities), $\hat{\boldsymbol{\rho}}_n(t)$ is the predicted spatiotemporal demand profile derived from the CA's global/historical perception scope, $\Delta\mathcal{K}_n(t)$ is the incremental knowledge update, and $\boldsymbol{\omega}_n(t)$ is the conflict resolution directive that resolves inter-agent disputes escalated via the uplink exception reports $\mathcal{E}_n(t)$.

        Unlike the traditional PCF$\to$gNB policy path, which disseminates static per-slice QoS profiles at administrative timescales, the CA$\to$NA directive conveys actionable intelligence that enables the NA to anticipate demand shifts, pre-configure resources, and adapt coordination strategies before performance degradation materializes. This proactive policy dissemination, informed by the CA's global knowledge base $\mathcal{K}(t)$ and cross-region trend analysis, directly addresses the reactive coordination bottleneck by equipping NAs with forward-looking context that is structurally unavailable in the traditional architecture.

        \paragraph{NA$\to$DA: Resource Grants and Model Updates}
        The NA responds to each DA with a Capability Abstraction Message (CAM):
        \begin{equation}
            \mathrm{CAM}_{n \to u}(t) = \bigl\{\, \hat{\mathbf{r}}_u(t),\; \boldsymbol{\alpha}_u(t),\; \Delta\mathbf{W}_u(t) \,\bigr\},
            \label{eq:cam}
        \end{equation}
        where $\hat{\mathbf{r}}_u(t)$ is the predicted resource availability profile, $\boldsymbol{\alpha}_u(t)$ is the resource grant confirming the allocated resources, and $\Delta\mathbf{W}_u(t)$ represents incremental model updates for the DA's local inference model.

        The CAM enables the DA to adapt its behavior proactively. This predictive capability, absent in traditional architectures where the UE has no visibility into future resource availability, is a direct consequence of the bidirectional semantic exchange enabled by the Agent Layer. The NA constructs the CAM by integrating the DA's uplink intent $\mathrm{SIM}_u(t)$ with the CA's downlink directive $\mathrm{Directive}_n(t)$, ensuring that per-user resource decisions are simultaneously aligned with individual user intent and global coordination policies.

        When the DA's intent cannot be fully satisfied by the NA's current resource allocation, a negotiation round is initiated. The DA revises its intent descriptor, and the NA proposes alternative resource configurations. This negotiation operates within a bounded number of rounds $N_{\mathrm{neg}}$ to ensure convergence within real-time constraints. The convergence guarantee is established through a cooperative game formulation where both agents optimize a shared utility function, ensuring that the negotiation converges to a Pareto-efficient outcome within the expanded solution space.

        \subsubsection{Peer-to-Peer Flow: NA$\leftrightarrow$NA}

        Adjacent Network Agents coordinate directly for inter-cell operations, including handover execution, inter-cell interference coordination (ICIC), and load balancing. This peer channel extends the traditional Xn interface signaling with richer semantic exchanges:
        \begin{equation}
            \mathrm{Peer}_{n \to m}(t) = \bigl\{\, \hat{\mathbf{L}}_n(t),\; \mathcal{U}_n^{\mathrm{HO}}(t),\; \boldsymbol{\xi}_{n,m}(t) \,\bigr\},
            \label{eq:peer_msg}
        \end{equation}
        where $\hat{\mathbf{L}}_n(t)$ is the predicted load profile enabling proactive load balancing, $\mathcal{U}_n^{\mathrm{HO}}(t)$ is the set of UEs predicted to require handover to cell~$m$ with associated intent descriptors and mobility predictions, and $\boldsymbol{\xi}_{n,m}(t)$ represents the interference coordination parameters negotiated between the two NAs.

\section{Operational Workflow}
\label{sec:workflow}

\begin{figure*}[!t]
    \centering
    \includegraphics[width=0.85\textwidth]{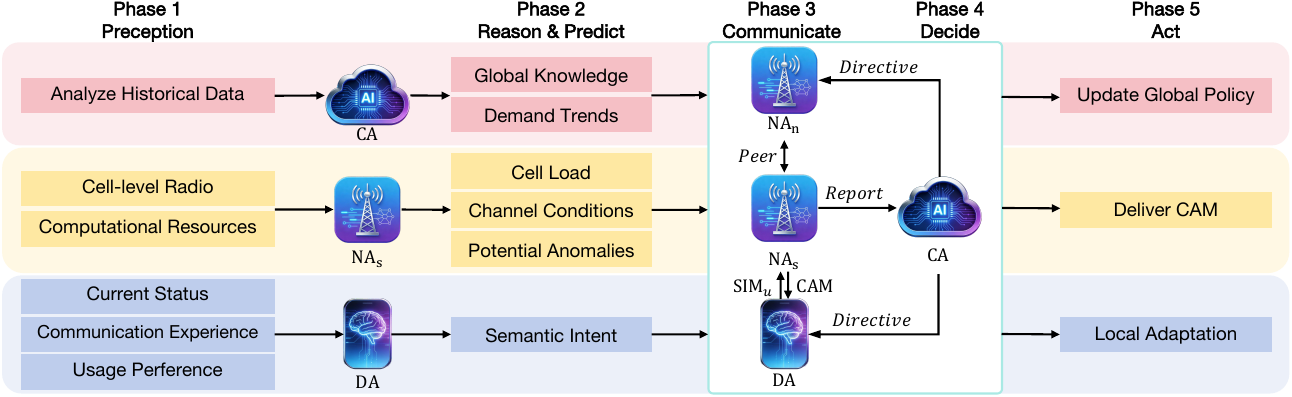}
    \caption{Collaborative operational workflow diagram in ASTRA application scenarios.}
    \label{fig:flow}
\end{figure*}

\begin{algorithm}[!t]
\caption{ASTRA Agentic Coordination Cycle}
\label{alg:astra}
\begin{algorithmic}[1]
\REQUIRE Task utility $U^{\mathrm{task}}(\cdot)$, cycle period $\Delta T$

\textbf{Phase 1: Perceive}
\STATE Each DA observes local state $\mathbf{s}_u^{\mathrm{DA}}(t)$; each NA observes cell state $\mathbf{s}_n^{\mathrm{NA}}(t)$; CA observes global state $\mathbf{s}^{\mathrm{CA}}(t)$

\textbf{Phase 2: Reason \& Predict}
\STATE DA infers intent $\boldsymbol{\phi}_u(t)$ and predicts QoE/mobility
\STATE NA predicts cell load $\hat{\mathbf{L}}_n(t{+}\Delta T)$, channel $\hat{\mathbf{h}}_n(t{+}\Delta T)$, and anomalies
\STATE CA predicts global demand $\hat{\boldsymbol{\rho}}(t{+}\Delta T)$ and detects cross-cell trends

\textbf{Phase 3: Communicate}
\STATE DA $\to$ NA: semantic intent message $\mathrm{SIM}_u(t)$
\STATE NA $\to$ CA: aggregated report $\mathrm{Report}_n(t)$
\STATE CA $\to$ NA: global directive $\mathrm{Directive}_n(t)$
\STATE NA $\leftrightarrow$ NA: peer coordination $\mathrm{Peer}_{n \to m}(t)$

\textbf{Phase 4: Decide}
\STATE Each DA solves $\mathbf{d}_u^{\star}(t) = \arg\max_{\mathbf{d}_u \in \mathcal{D}_u} U_{\mathrm{DA}}^{\mathrm{task}}(\mathbf{d}_u \mid \mathbf{s}_u^{\mathrm{DA}}, \mathrm{CAM}_{n \to u})$
\STATE Each NA solves $\mathbf{a}_n^{\star}(t) = \arg\max_{\mathbf{a}_n \in \mathcal{X}_n} U_{\mathrm{NA}}^{\mathrm{task}}(\mathbf{a}_n \mid \mathbf{s}_n^{\mathrm{NA}}, \mathrm{Directive}_n, \{\mathrm{Peer}_{m \to n}\})$
\STATE CA solves $\boldsymbol{\Pi}^{\star}(t) = \arg\max_{\boldsymbol{\Pi}} U_{\mathrm{CA}}^{\mathrm{task}}(\boldsymbol{\Pi} \mid \mathbf{s}^{\mathrm{CA}}, \{\mathrm{Report}_n\})$
\STATE On conflict: bounded negotiation ($\leq N_{\mathrm{neg}}$ rounds) until Pareto agreement

\textbf{Phase 5: Act}
\STATE DA executes $\mathbf{d}_u^{\star}(t)$ locally; NA executes $\mathbf{a}_n^{\star}(t)$ on Physical Layer and sends $\mathrm{CAM}_{n \to u}(t)$; CA publishes $\boldsymbol{\Pi}^{\star}(t)$

\textbf{Phase 6: Learn}
\STATE DA updates $\mathbf{z}_u$; NA updates $\boldsymbol{\psi}_n$; CA updates $\mathcal{K}$ based on decisions and outcomes
\end{algorithmic}
\end{algorithm}

    \subsection{ASTRA Coordination Algorithm}

    The ASTRA paradigm executes a six-phase cycle across the DA, NA, and CA tiers, as summarized in \textbf{Algorithm~\ref{alg:astra}}. Each cycle is triggered either periodically at a task-dependent cadence $\Delta T$ or on-demand when any agent detects a significant state transition (e.g., QoE degradation, load surge, or mobility event).

    The cycle directly addresses the three structural bottlenecks identified in Section~\ref{sec:tradition}: (i)~Phase~4 optimizes over the full decision space $\mathcal{X}$ instead of the protocol-constrained $\mathcal{X}^{\mathrm{3GPP}}$; (ii)~Phase~3 retains semantic context through bidirectional semantic channels, mitigating the cumulative mutual information loss $\mathcal{L}_{\mathrm{total}}$ inherent in cascaded traditional interfaces; and (iii)~Phase~2 enables proactive coordination via predictive models, eliminating the reactive latency $T_{\mathrm{degrade}} + T_{\mathrm{TTT}}$ that pervades the traditional architecture.

    \subsection{Application Scenarios and Case Analysis }

    By specializing $U^{\mathrm{task}}(\cdot)$ and agent decision spaces, the generic six-phase cycle instantiates a broad class of DNC coordination tasks. The ASTRA paradigm is designed as a general-purpose coordination framework: any operational task that can be formulated as optimizing a utility function over distributed observations, where agents possess heterogeneous perception scopes and can exchange semantic intent, is in principle amenable to the agentic coordination cycle. Fig.~\ref{fig:flow} illustrates the collaborative operational workflow. Below, we first provide a structured taxonomy of application domains that the paradigm supports, and then present two concrete cases validated through system-level simulations.

        \paragraph{Taxonomy of Supported Application Domains}
        The ASTRA paradigm exhibits broad applicability across multiple dimensions of DNC operations. We categorize representative application domains into four classes, each corresponding to a distinct specialization of the agent decision spaces and utility functions in Algorithm~\ref{alg:astra}:

        \begin{itemize}
        \item \textbf{Radio Resource Management (RRM).} Beyond cell selection, the joint optimization over $\mathcal{X}_{\mathrm{radio}}$ informed by semantic intent extends to intent-aware spectrum allocation, dynamic power control, multi-user MIMO precoding where beamforming weights incorporate predicted UE mobility and QoE trajectories, and coordinated multi-point (CoMP) transmission where multiple NAs negotiate joint transmission configurations based on shared UE intent rather than static cluster definitions. The key enabler is the replacement of aggregate, backward-looking KPI proxies with forward-looking, per-UE semantic intent as the optimization objective.

        \item \textbf{Mobility Management.} Beyond high-speed handover, the proactive mobility framework extends to UAV trajectory-aware cell association where the DA predicts its 3D flight path and negotiates sequential handovers along the route before takeoff, satellite-terrestrial handover in non-terrestrial network scenarios where the DA predicts satellite pass durations and pre-schedules link switches, and heterogeneous network tier selection where the DA jointly considers small-cell and macro-cell coverage, backhaul capacity, and application latency tolerance to select the optimal serving tier.

        \item \textbf{Computation-Centric Orchestration.} The semantic task graph $\mathbf{g}_u(t)$ carried by the SIM enables computation-aware coordination including model partitioning for on-device/edge/cloud split inference, federated learning task scheduling where the CA selects participating DAs based on data quality and channel conditions encoded in intent descriptors, and edge caching where the NA predicts content popularity from intent patterns and pre-caches segments before the DA requests them.

        \item \textbf{Service-Level Orchestration.} The semantic intent channel enables coordination above the radio layer: dynamic network slicing where the NA reconfigures slice resource partitions in real time based on the aggregated intent criticality distribution across served UEs; QoS-to-intent mapping where the CA learns a mapping from application-layer intent descriptors to optimal 5QI assignments; and intent-driven traffic steering where the NA routes UE traffic across multiple user-plane paths based on per-flow latency and throughput intent.
        \end{itemize}

        Beyond these operational domains, the ASTRA paradigm further provides cross-cutting capabilities including cross-layer anomaly detection, root-cause analysis, self-healing, and predictive maintenance (network resilience), as well as generic multi-agent primitives such as auction-based resource allocation, coalition formation, and hierarchical task decomposition that compose into more complex workflows. The breadth of applicability arises from a single architectural invariant: the ASTRA cycle in Algorithm~\ref{alg:astra} is task-agnostic---it does not hard-code any specific optimization objective, radio access technology, or deployment topology. Specialization requires only: (i) defining the task-specific utility function $U^{\mathrm{task}}(\cdot)$, (ii) instantiating the agent state vectors with domain-relevant observations, and (iii) configuring the semantic message formats with task-appropriate descriptors. The paradigm defines how agents collaborate, not what they collaborate about.

        Below, we elaborate two representative cases that have been validated through system-level simulations in Section~\ref{sec:performance}.

        \paragraph{Case~1: Dense-Crowd Cell Selection.}
        In large-scale public events, users concentrate within overlapping cell coverage. The standard 3GPP criterion $n^{\star}_{\mathrm{3GPP}} = \arg\max_{n} \mathrm{RSRP}_n$ drives all nearby UEs toward the strongest cell~\cite{3gpp_ts_38304}, causing severe congestion while neighboring cells remain underutilized. In ASTRA, each DA continuously monitors its per-user throughput and QoE. Upon detecting throughput degradation despite adequate RSRP, the DA autonomously infers that the root cause is cell overloading---a causal reasoning step beyond the capability of traditional measurement reports---and encodes this congestion experience, together with its application criticality and desired QoS target, into a Semantic Intent Message (SIM) transmitted uplink to its serving NA. Meanwhile, the serving NA broadcasts load queries to neighboring NAs via the peer channel ($\mathrm{Peer}_{n \to m}$), soliciting their available capacity profiles. Armed with per-UE congestion intents from DAs, neighbor capacity offers, and global load-balancing directives from the CA, the NA autonomously determines which UEs to offload and to which target cells. In the proposed paradigm, the NA is designed to reason over the semantic content of each SIM rather than following fixed heuristics; in our proof-of-concept implementation, this reasoning is realized through LLM-based inference. Once consensus is reached, the NA delivers a CAM to each reassigned DA confirming the new serving cell, executes the handover, and the CA subsequently aggregates offloading outcomes to refine the global load-balancing policy. The process converts the herding behavior of max-RSRP into globally-aware, intent-prioritized load balancing, where the coordination logic emerges from semantic exchange and autonomous reasoning rather than a closed-form utility.

        \paragraph{Case~2: High-Speed Mobility Optimization.}
        In a high-speed corridor, the bottleneck shifts from spatial load imbalance to temporal urgency: the UE traverses each cell in seconds, and reactive A3-event handovers incur $T_{\mathrm{degrade}} + T_{\mathrm{TTT}}$ of degradation before the trigger fires. In ASTRA, the DA fuses onboard sensors (GPS, IMU, speedometer) to construct a short-term trajectory prediction $\hat{\mathbf{p}}_u(t+\Delta t)$ and estimates its residual dwell time, then transmits a SIM carrying this mobility forecast together with application continuity requirements (e.g., the maximal tolerable interruption for streaming versus the zero-outage mandate for autonomous driving). The serving NA forwards the UE's trajectory prediction to candidate target NAs along the projected path via the peer channel, requesting advance resource reservation---a pre-negotiation that occurs before the UE reaches the cell edge, in contrast to the A3-event architecture where the target cell is contacted only after the trigger fires. Each candidate NA responds with its predicted resource availability at the projected arrival time. The serving NA then weighs these responses against the UE's continuity requirements: for a streaming UE it may defer handover to avoid ping-pong; for an autonomous-driving UE it triggers immediate handover with confirmed resource reservation. The NA delivers a CAM specifying the target cell and expected handover time, then initiates the procedure before the UE enters the degradation zone, eliminating $T_{\mathrm{degrade}} + T_{\mathrm{TTT}}$ from the latency budget. The CA subsequently aggregates handover outcomes across all NAs along the corridor to refine the global mobility policy, and the DA updates its trajectory prediction model from the forecast accuracy. Unlike a fixed mobility-weighted formula, this process enables context-aware handover where the decision adapts to application semantics, real-time network state, and multi-agent negotiation outcomes.

\section{Performance Evaluation}
\label{sec:performance}

To validate ASTRA, we conduct system-level simulations for both cases described in Section~\ref{sec:workflow} using the San Francisco DeepMIMO ray-tracing dataset~\cite{alkhateeb2019deepmimo}. Lightweight LLMs are deployed across the three agent tiers to enable autonomous reasoning: Qwen-3.5-2B on the device side, Qwen-3.5-9B at the network side, and Qwen-3.5-122B in the cloud. The baseline for all comparisons is the standard 3GPP protocol-driven architecture (Max-RSRP cell selection, A3-event handover with static hysteresis).

\subsection{Dense-Crowd Cell Selection}

The dense-crowd scenario deploys 5 base stations in an asymmetric layout at 3.5~GHz with 100~MHz bandwidth, covering a $318 \times 268$~m urban microcell environment with 3300~UEs distributed non-uniformly around a central hotspot, emulating a crowded public venue.

\begin{figure}[!t]
    \begin{minipage}{0.47\textwidth}
        \centering
        \includegraphics[width=\linewidth]{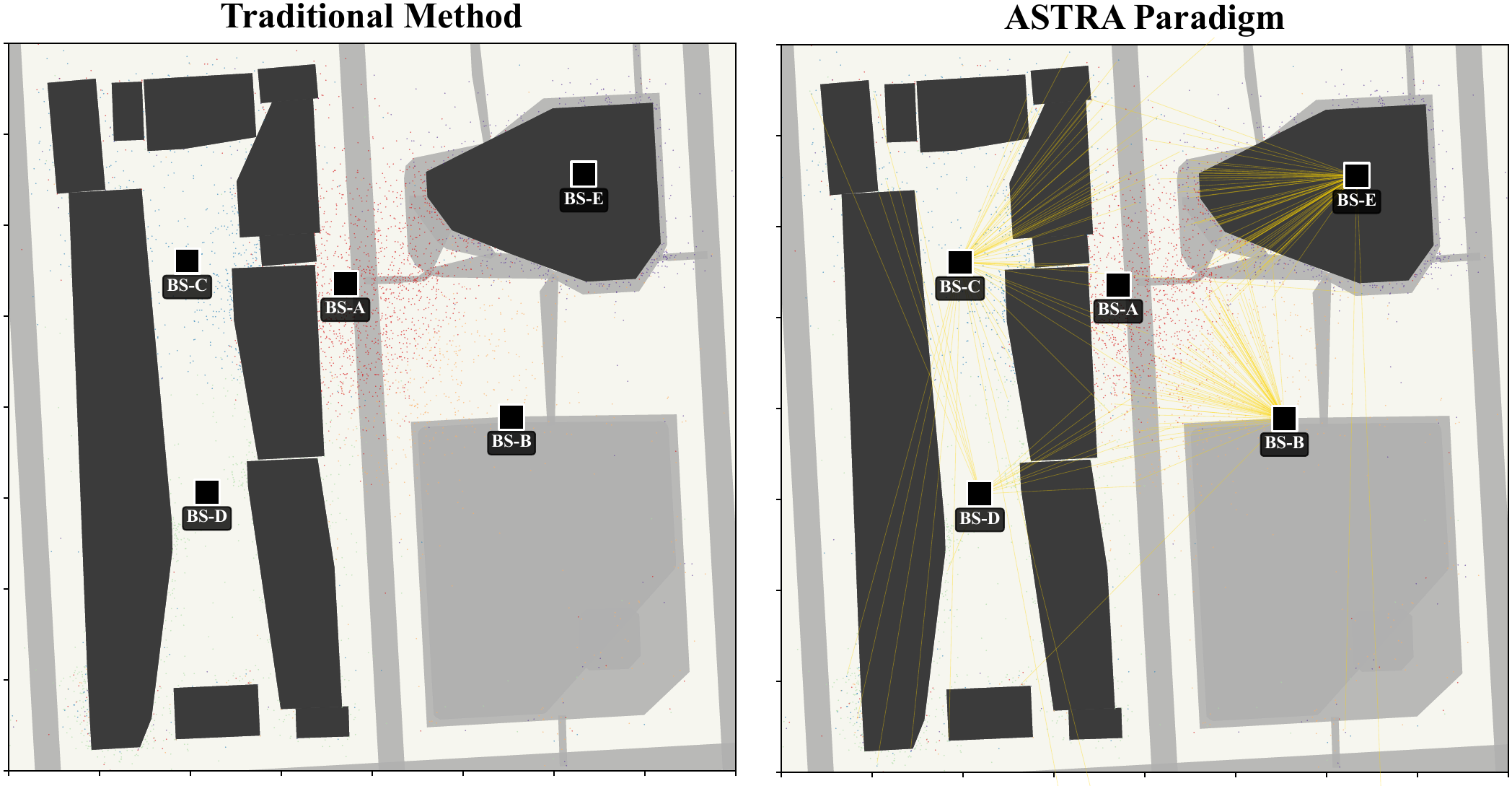}
        \caption{UE-cell association map under baseline Max-RSRP and ASTRA. ASTRA redistributes 203~UEs (6.2\%) from the congested central cell to neighboring cells via NA$\leftrightarrow$NA semantic load exchange.}
        \label{fig:topology}
    \end{minipage}
    \hfill
    \begin{minipage}{0.47\textwidth}
        \centering
        \includegraphics[width=0.8\linewidth]{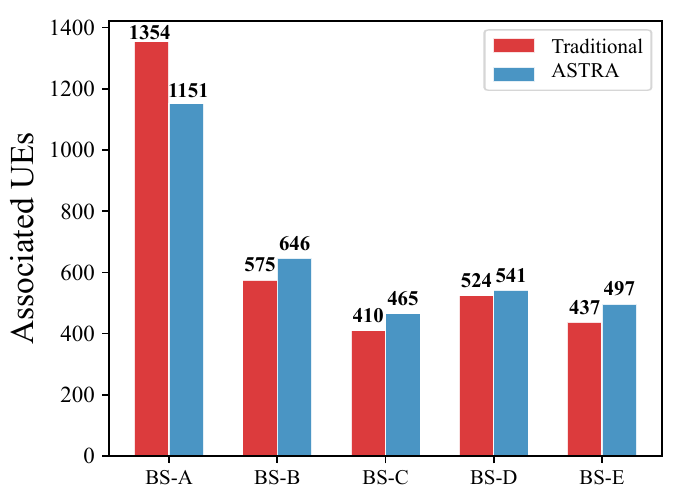}
        \caption{UE-cell association under ASTRA with per-cell load distribution annotated.}
        \label{fig:cell_crowd}
    \end{minipage}
    \vspace{-0.5em}  
\end{figure}

Fig.~\ref{fig:topology} visualizes the UE-cell association before and after ASTRA coordination. Under Max-RSRP, the central BS attracts 1354~UEs (41.0\%), creating severe congestion while peripheral cells remain underutilized. ASTRA resolves this imbalance through agent collaboration: congested DAs autonomously infer overload from the discrepancy between strong RSRP and poor throughput, encode this experience into Semantic Intent Messages, and report uplink to the serving NA. The NA exchanges load information with neighboring NAs via the peer channel and, guided by global directives from the CA, negotiates which UEs to offload and to which target cells. This semantic load exchange reassociates 203~UEs (6.2\%) to neighboring cells, reducing the load standard deviation from 352.0 to 253.0 ($-28.1\%$). Fig.~\ref{fig:cell_crowd} shows the resulting ASTRA association with per-cell load annotated, confirming that the redistribution is not a centralized heuristic but an emergent outcome of autonomous multi-agent reasoning.

\begin{figure}[!t]
    \centering
    \includegraphics[width=0.4\textwidth]{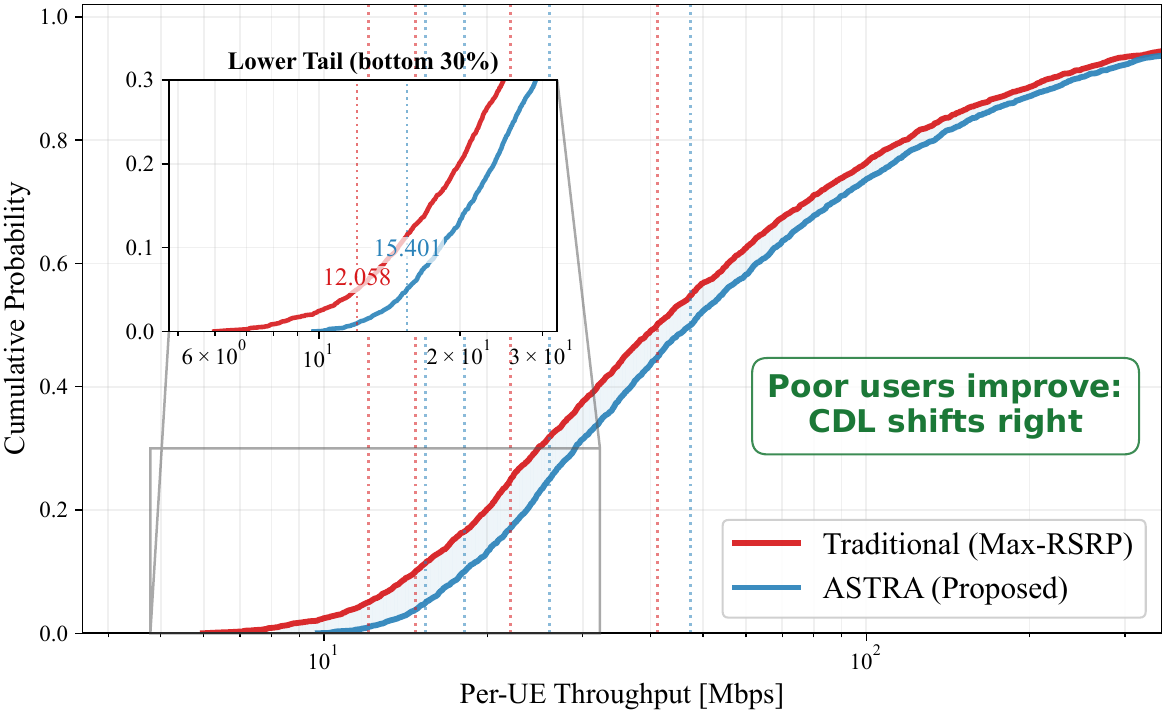}
    \caption{Per-UE throughput CDF under baseline and ASTRA. The ASTRA curve is systematically right-shifted across the entire distribution.}
    \label{fig:throughput_percentile}
\end{figure}

Fig.~\ref{fig:throughput_percentile} presents the per-UE throughput CDF. The ASTRA curve is systematically right-shifted across the entire throughput range, indicating that agentic coordination improves experience uniformly rather than trading one UE group against another. Quantitatively, the mean throughput rises from 85.98 to 97.28~Mbps ($+13.1\%$), the median from 41.13 to 47.48~Mbps ($+15.4\%$), and the 5th-percentile---a critical metric for cell-edge users---from 12.06 to 15.40~Mbps ($+27.7\%$). The disproportionately large gain at the low percentile confirms that ASTRA's semantic intent exchange converts the herding behavior of max-RSRP into globally-aware, intent-prioritized load balancing, producing a double dividend: offloaded UEs benefit from the target cell's spare capacity, while remaining UEs benefit from reduced contention. The mean MOS improves from 4.74 to 4.77, and the Jain fairness index increases from 0.372 to 0.378.

\subsection{High-Speed Mobility Optimization}

The high-speed scenario simulates a 3.0~km highway corridor with 5 base stations deployed alternately along both sides of the road at 3.5~GHz. A total of 300 vehicles travel at speeds uniformly distributed in 81--298~km/h. The simulation runs for 20~s at a 0.1~s time step (200 slots), with each handover incurring a 150~ms effective outage.

\begin{figure}[!t]
    \centering
    \includegraphics[width=0.48\textwidth]{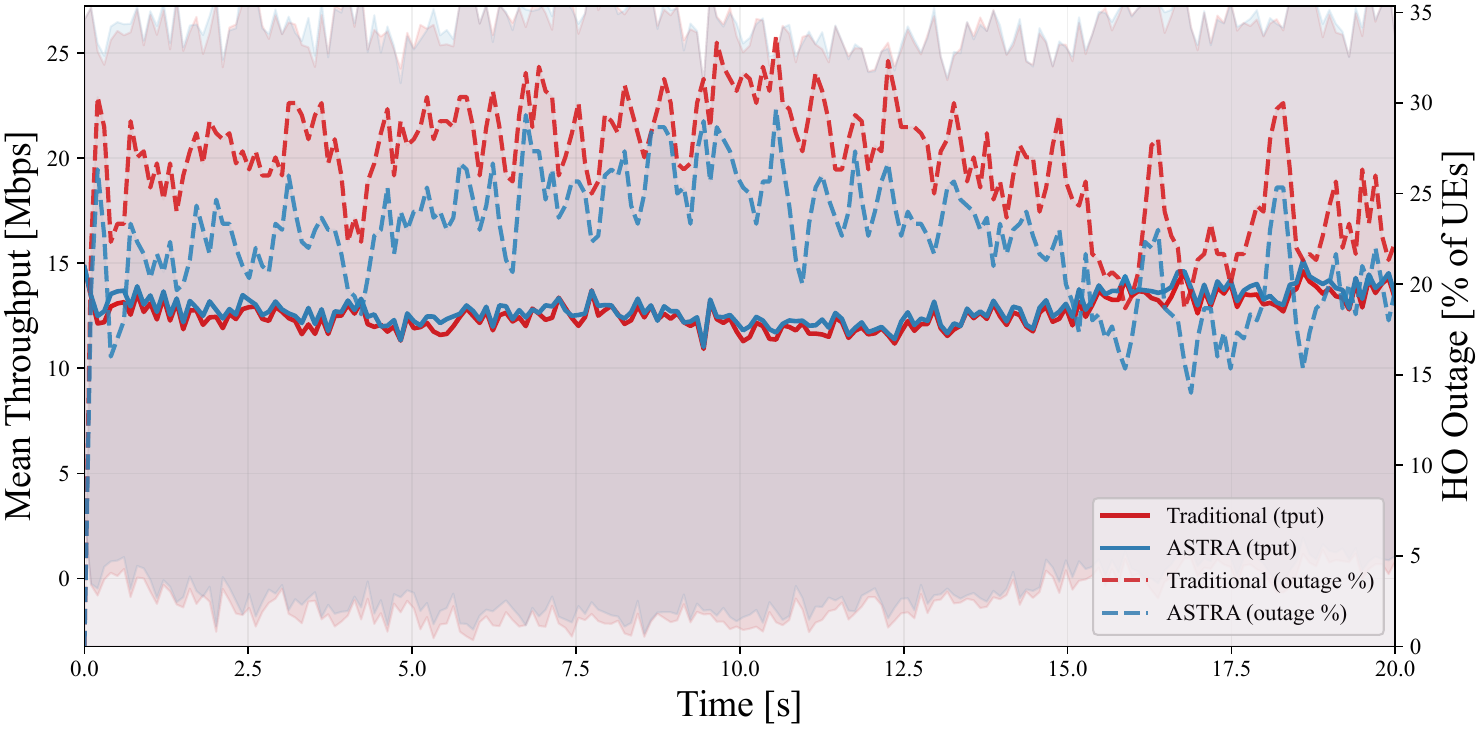}
    \caption{Temporal dynamics of per-UE mean throughput (solid) and instantaneous handover outage ratio (dashed) under baseline vs.\ ASTRA.}
    \label{fig:throughput_outage}
\end{figure}

Fig.~\ref{fig:throughput_outage} presents the time-series evolution on dual vertical axes. Solid curves track per-UE mean throughput; dashed curves track the instantaneous fraction of vehicles experiencing zero throughput due to handover interruption. Under the baseline reactive A3-event scheme, passive handovers occur at 1.84~HO/s with a mean throughput of 12.60~Mbps and a stability index of 0.815. ASTRA reduces the handover rate to 1.50~HO/s ($-18.2\%$), a direct consequence of the predictive coordination chain: the DA forecasts vehicle trajectory from onboard sensors, the serving NA pre-negotiates resource reservation with candidate target NAs via the peer channel before the UE reaches the cell edge, and the NA initiates handover only when the target cell confirms resource availability---eliminating $T_{\mathrm{degrade}} + T_{\mathrm{TTT}}$ from the latency budget. The mean throughput rises to 12.88~Mbps ($+2.2\%$) and stability to 0.849, with the Jain fairness index improving from 0.823 to 0.826. The modest throughput gain is consistent with the nature of this scenario: the primary bottleneck is not capacity but handover-induced outage, and ASTRA's principal contribution is converting reactive, loss-prone handover into a proactive, intent-aware process. The simultaneous improvement in throughput, stability, and fairness confirms that predictive coordination resolves the structural trade-off imposed by the conventional A3-event mechanism.

\section{Conclusion}
\label{sec:conclusion}

This paper proposed ASTRA, an autonomous agentic AI paradigm that envisions a restructuring of the DNC continuum for next-generation mobile networks by introducing a three-tier agent layer (DA, NA, CA), bidirectional semantic channels, and a six-phase proactive coordination cycle. System-level simulations in two representative 6G scenarios demonstrate that the proposed framework accesses solution regions structurally inaccessible under the prevailing protocol-constrained architecture: a 13.1\% average throughput gain and 27.7\% cell-edge throughput improvement in dense-crowd cell selection through semantic load-aware UE redistribution, and an 18.2\% passive handover reduction in high-speed mobility through predictive trajectory-aware coordination.

Several limitations of the current study warrant discussion. First, the validation is conducted exclusively through system-level simulations; real-world deployment on operational 5G infrastructure would be necessary to characterize implementation overheads, including LLM inference latency at each agent tier, inter-agent communication bandwidth, and integration complexity with existing protocol stacks. Second, the current evaluation considers two canonical scenarios; the paradigm's performance under more heterogeneous deployments involving satellite-terrestrial integration, UAV-assisted networks, and industrial IoT environments remains an open question. Third, security and privacy aspects of exposing semantic user intent to network and cloud agents require formal treatment, particularly regarding intent data confidentiality, agent authentication, and adversarial robustness of LLM-based decision-making.

Future work will pursue three directions: (i) prototype implementation on O-RAN testbeds with open-source LLM backends to quantify real-time feasibility and latency characteristics of the agentic cycle; (ii) extension to multi-vendor and multi-operator scenarios where agents negotiate across administrative boundaries, requiring standardization of the proposed semantic message formats; and (iii) integration of foundation-model-based agents with classical optimization solvers for safety-critical decisions where formal optimality guarantees are mandated. More broadly, the ASTRA paradigm opens a research avenue toward self-evolving network intelligence where the coordination protocol itself is subject to continuous learning and adaptation.

\bibliographystyle{IEEEtran}
\bibliography{Intro}

\vfill

\end{document}